\documentclass[english,twocolumn]{revtex4-1}
\usepackage[T1]{fontenc}
\usepackage[latin9]{inputenc}
\usepackage{verbatim}
\usepackage{graphicx}


\usepackage{babel}
\begin{document}

\title{Metastable Carbon at Extreme Conditions}

\author{Ashley S. Williams}

\affiliation{Department of Physics, University of South Florida, Tampa, FL 33620}

\author{Kien Nguyen-Cong}

\affiliation{Department of Physics, University of South Florida, Tampa, FL 33620}

\author{Jonathan T. Willman}

\affiliation{Department of Physics, University of South Florida, Tampa, FL 33620}

\author{Ivan I. Oleynik}
\email{oleynik@usf.edu}

\selectlanguage{english}%

\affiliation{Department of Physics, University of South Florida, Tampa, FL 33620}
\begin{abstract}
Carbon at extreme conditions is the focus of intensive scientific
inquiry due to its importance for applications in inertial confinement
fusion experiments and for understanding the interior structure of
carbon-rich exoplanets. The extreme metastability of diamond at very
high pressures has been discovered in recent dynamic compression experiments.
This work addresses an important question about the existence of other
competitive metastable carbon phases that might be observed in shock
experiments. It was found that diamond polytypes, carbon crystals
with mixed cubic and hexagonal diamond stacking planes, are the only
metastable carbon crystal phases energetically competitive with cubic
diamond at pressures between 100 and 1,000 GPa. Above 1 TPa, no metastable
phases are found to be energetically competitive with thermodynamically
stable BC8 and simple cubic phases. The existence of low enthalpy
diamond polytypes suggests that they are likely candidates for metastable
phases of carbon to appear upon shockwave loading of diamond .
\end{abstract}
\maketitle
Carbon, one of the most frequently occurring elements in the universe,
is unique due to its ability to form sp, sp$^{2}$, and sp$^{3}$
hybrid orbitals resulting in numerous metastable allotropes at ambient
conditions in addition to thermodynamically stable graphite and slightly
metastable cubic diamond (CD). Although carbon is well studied at
ambient conditions its behavior at extreme conditions is not well
understood. Recently, the high pressure behavior of carbon has attracted
substantial effort, both experimentally \citep{Kondo1983,Scandolo1995,Knudson2008,Eggert2009,McWilliams2010,Kraus2012,Smith2014,Jones2016,Nemeth2020,Nemeth2020a,Turneaure2017,Stavrou2020,Volz2020}
and theoretically \citep{Correa2008,Oleynik2008,Wen2008,Sun2009,Zhu2011,Martinez-Canales2012,Oganov2013,Pineau2013,Cui2015,Mujica2015,Li2018,Zhu2020}.

Not much is known about the metastability of carbon under extreme
conditions and the topic has become somewhat controversial as new
experimental and simulation methods become prevalent. For instance,
while Knudson \textit{et al.}\citep{Knudson2008} showed the appearance
of BC8 phase near the triple point, recent dynamic compression at
NIF\citep{Smith2014} failed to observe transition of CD to BC8 and
instead suggested a high energy barrier between the two phases. The
notion of a high energy barrier lead to the proposal of metastability
of diamond extending well beyond its region of thermodynamic stability,
which begs the question- are there metastable phases at extreme conditions
that are being missed with conventional methods? Along the same line,
x-ray diffraction experiments have claimed to observe hexagonal diamond
from shock loading of both graphite \citep{Turneaure2017,Stavrou2020,Volz2020,Dong2020}
and diamond \citep{Baek2019,He2002,Jones2016,Murri2019}, however,
there are still questions surrounding these observations. One major
question being posed is whether or not it is possible to resolve such
smaller differences in the crystals and definitively say when transformation
had occurred. This work aims to answer these questions with a systematic
study of carbon up to 5 TPa under uniaxial shock compression conditions
as this is the only way to experimentally reach such high pressures
in carbon. We find that carbon forms few energetically competitive
metastable structures at high pressures other than the known thermodynamically
stable phases of graphite, CD, BC8, and simple cubic (SC). Diamond
polytypes, a class of carbon crystal that contains mixing of CD stacking
and hexagonal diamond (HD) stacking, were the only competitive phases
possible in pressures up to 1 TPa. In agreement with previous investigations
\citep{Correa2008,Martinez-Canales2012}, there were no phases found
that could compete with thermodynamically stable BC8 or SC above 1TPa.

\begin{figure}[b]
\includegraphics[scale=0.25]{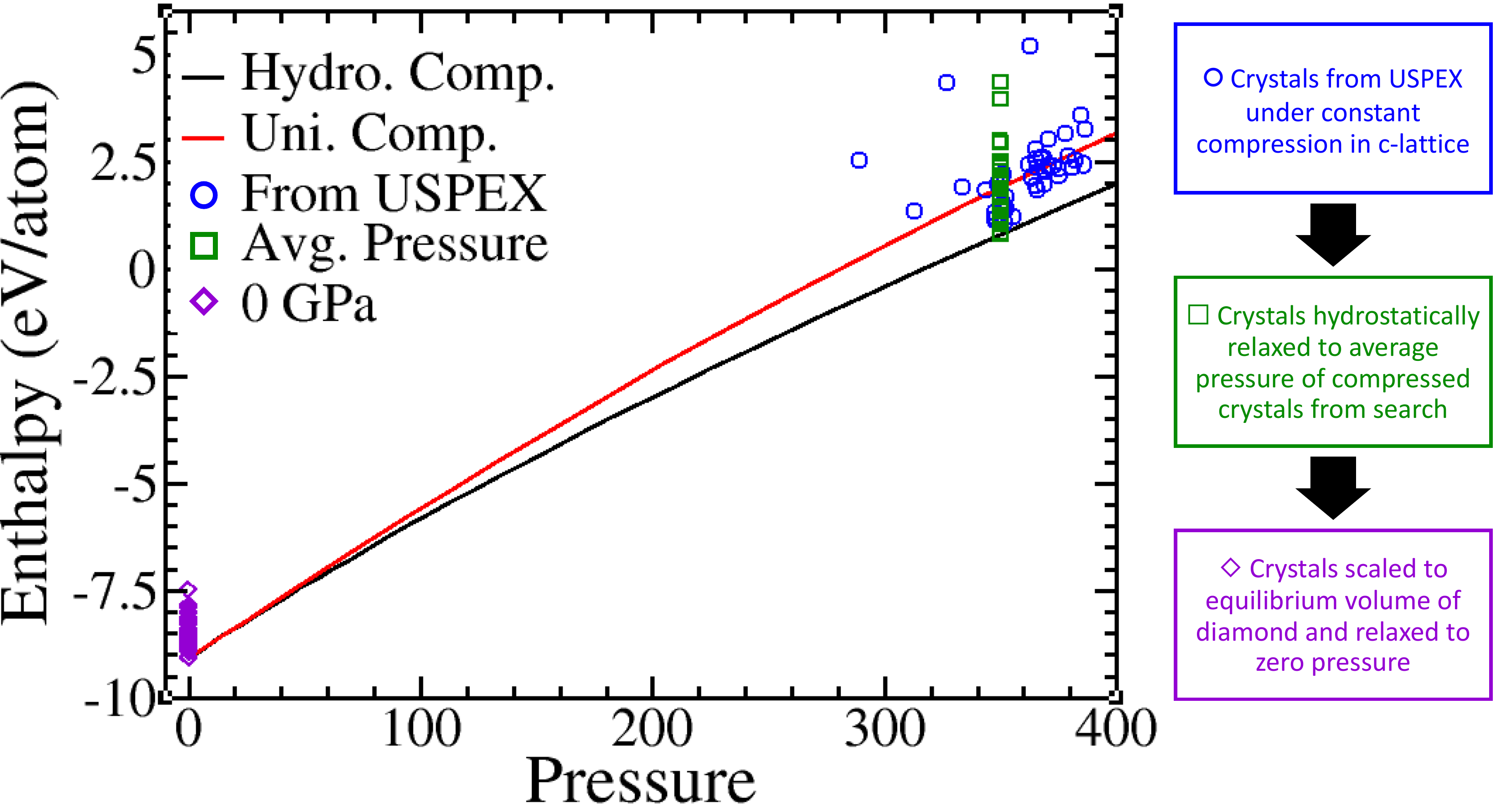}

\caption{\label{fig:Uni-Schematic}Schematic showing procedure for analysis
of uniaxially compressed USPEX searches. Black line corresponds to
the hydrostatic compression of diamond. Red line corresponds to uniaxial
compression of <100> diamond. Structures originated from search at
fixed lattice corresponding to <100> diamond with longitudinal stress
of 500 GPa.}
\end{figure}

In order to simulate shock wave conditions, we extend the crystal
structure prediction method \citep{Glass2006,Oganov2006,Lyakhov2010}
beyond hydrostatic conditions by applying uniaxial compression to
the unit cell of diamond. The evolutionary algorithm in the crystal
structure prediction method works by ranking individuals via enthalpy
which is not uniquely defined under uniaxial compression. Therefore,
careful consideration of the energetics of predicted crystals that
mimics shock wave experiments must take place. First, rapid uniaxial
compression takes places via the search followed by relaxation to
hydrostatic conditions at a given pressure then final release to ambient
conditions. All individuals from the final generation of the search
are considered in order to get accurate rankings upon the two instances
of hydrostatic relaxation. The protocol is detailed in Fig. \ref{fig:Uni-Schematic}.

\begin{figure*}
\includegraphics[scale=0.5]{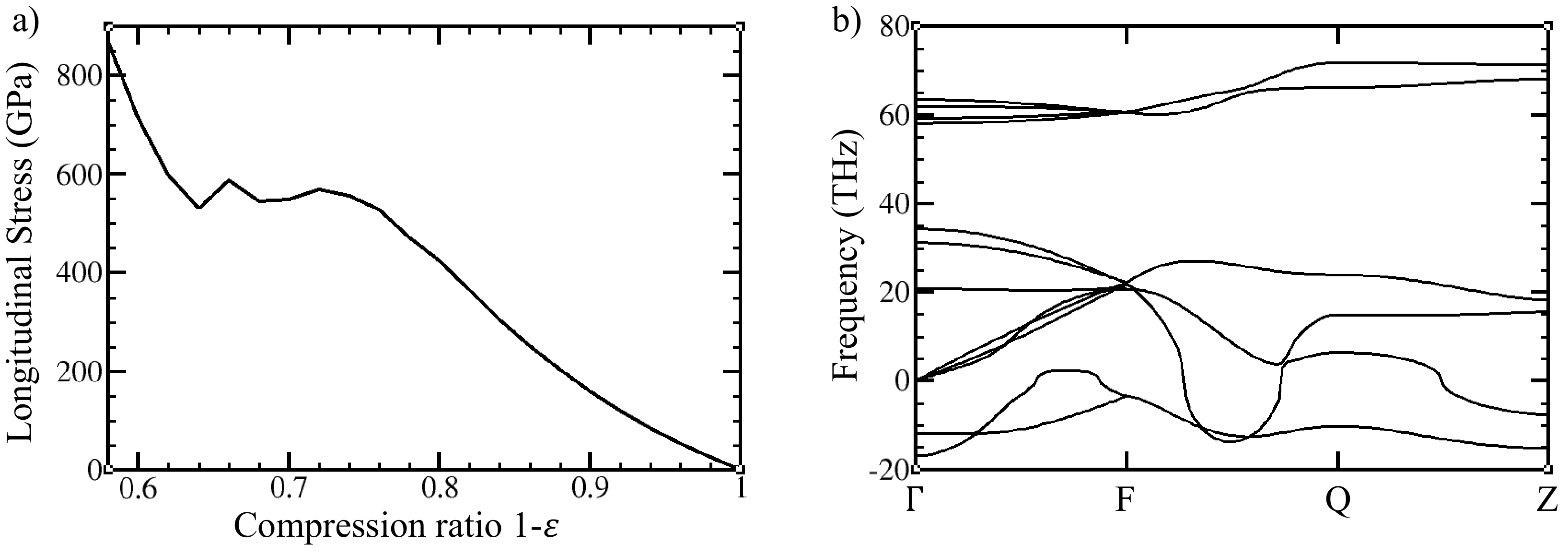}

\caption{\label{fig:110-Instability}a) Stress versus strain curve for uniaxially
compressed <110> diamond with change in slope around 500 GPa suggesting
instability of the crystal beyond this point. b) Calculated phonon
spectra of <110> diamond uniaxially compressed to 500 GPa longitudinal
stress with negative frequencies. }
\end{figure*}

\begin{figure}
\includegraphics[scale=0.3]{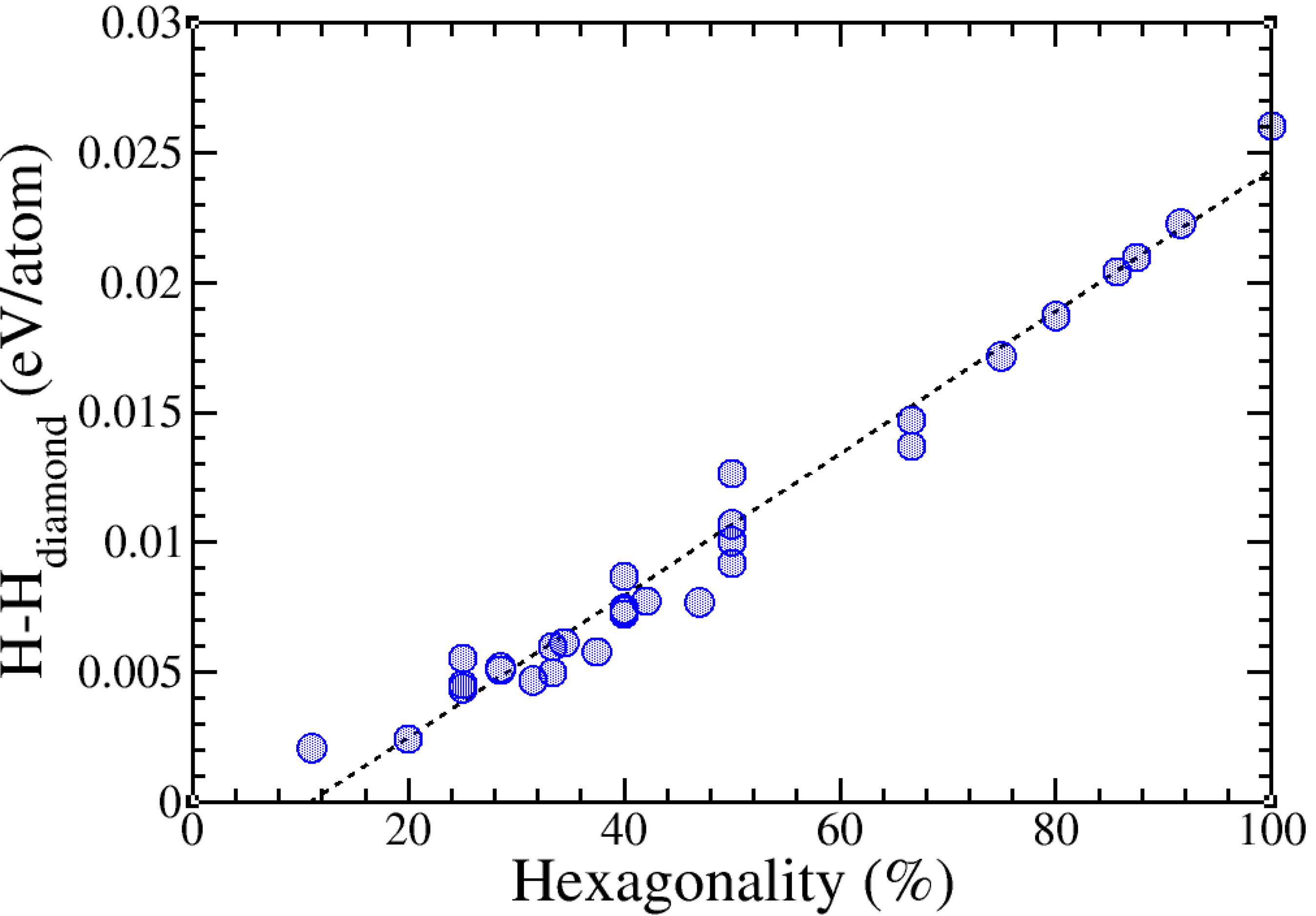}

\caption{\label{fig:Hexagonality-Energetics}Relative enthalpy of a sample
of diamond polytypes as a function of hexagonality showing an increasing
trend from fully CD stacked crystal to fully HD stacked crystal. Crystal
structure for polytypes not found in our searches or built by hand
taken from databases \citep{Jain2013,Grazulis2009}.}

\end{figure}

\begin{figure}
\includegraphics[scale=0.3]{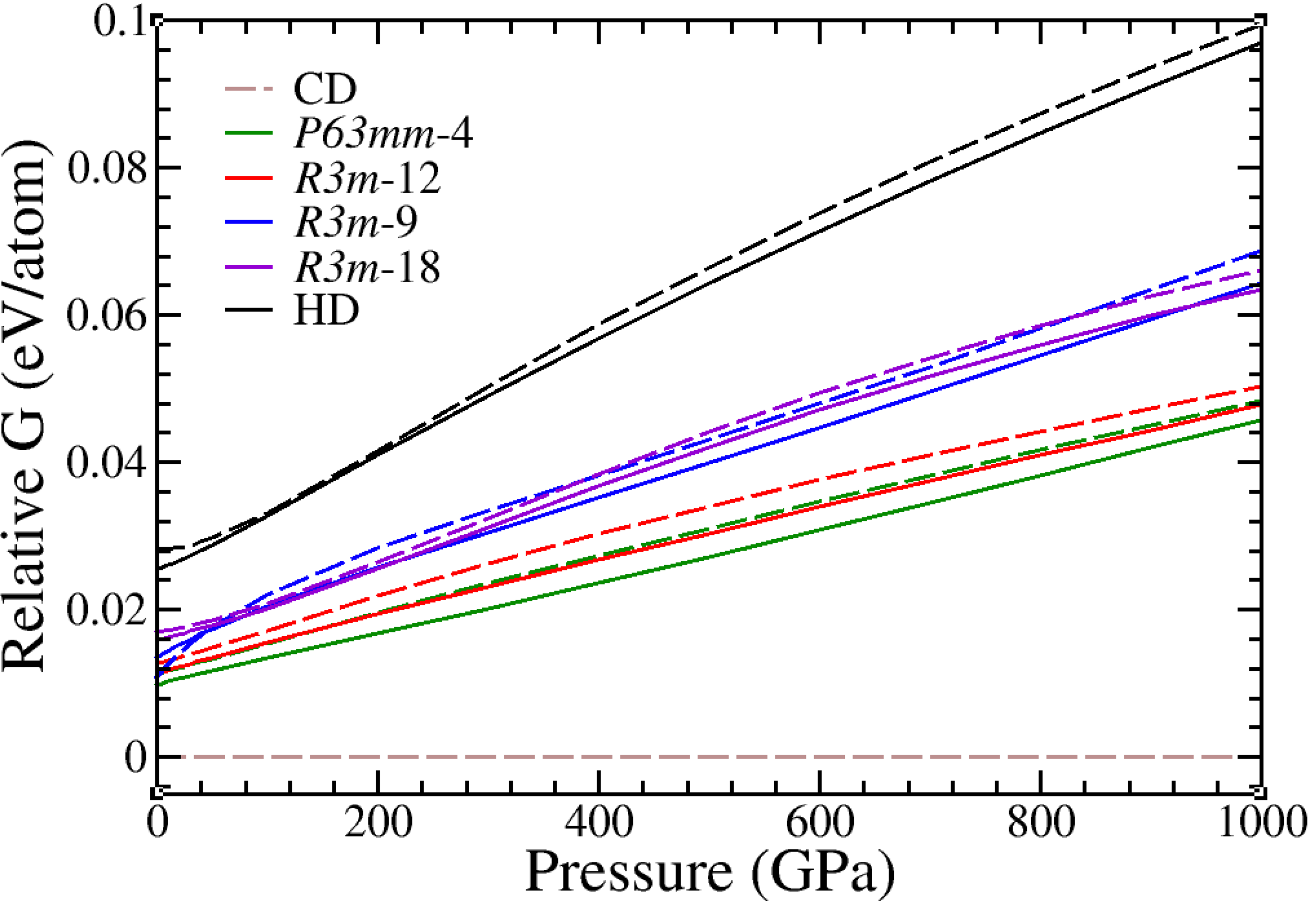}

\caption{\label{fig:Temp-Dependence}Temperature dependence of enthalpy for
selected polytypes showing no significant changes.}
\end{figure}

We used this protocol to confirm the previously known hydrostatic
phase diagram of carbon \citep{Martinez-Canales2012,Oganov2013} up
to 5 TPa via several first-principles crystal structure searches at
pressures ranging from 100 to 5000 GPa. All predicted crystals within
0.1 eV/atom of the lowest enthalpy crystal from the final generation
of the search were considered for the hydrostatic phase diagram. While
the story of hydrostatically compressed carbon is a simple one to
theoretically explore and tell, its behavior under uniaxial compression
is not. Investigation of the stress-strain curve of uniaxially compressed
<110> diamond shows non-monotonic behavior suggesting instability
in the crystal, see Fig. \ref{fig:110-Instability}a. Phonon band
structure of the <110> diamond uniaxially compressed to 500 GPa longitudinal
stress, Fig. \ref{fig:110-Instability}b, shows negative frequencies
indicating there is indeed dynamic instability of the crystal under
uniaxial compression. Searches at 100 GPa longitudinal stress predict
conventional uniaxially compressed diamond to be the preferred orientation.
However, we predict rearrangement of the atoms inside the unit cell
for most searches performed under uniaxial compression at longitudinal
stresses greater than 100 GPa.%
\begin{comment}
I originally had that the searches found different crystals to be
energetically favorable at 300 GPa and above but I took that out as
I am nervous to make such a strong conclusion about the energetics
of the searches.
\end{comment}
{} The searches predict that diamond polytypes, i.e. crystals that contain
a mixture of CD and HD stacking, are the most likely metastable phases
able to be achieved during shock experiments up to 1 TPa. These results
suggest that shock compression of diamond beyond 100 GPa longitudinal
stress is necessary to observe a transformation into these predicted
phases. Despite the suggestion\citep{He2002} that HD is formed via
shock compression of diamond, our results predict that diamond polytype
crystals with lower hexagonality are energetically preferred as seen
in Figure \ref{fig:Hexagonality-Energetics}. Here we define the hexagonality
of a crystal as the ratio of the number of HD layers to the total
number of layers; with a layer being defined as two consecutive atoms
in the same c-lattice plane. For naming convention, we will refer
to all polytypes by their symmetry and number of layers. As a rise
in temperature is inherently associated with shock compression, we
provide the temperature dependance of the enthalpies of a sample of
polytypes along with HD in Figure S\ref{fig:Temp-Dependence}.

\begin{figure*}
\includegraphics[scale=0.35]{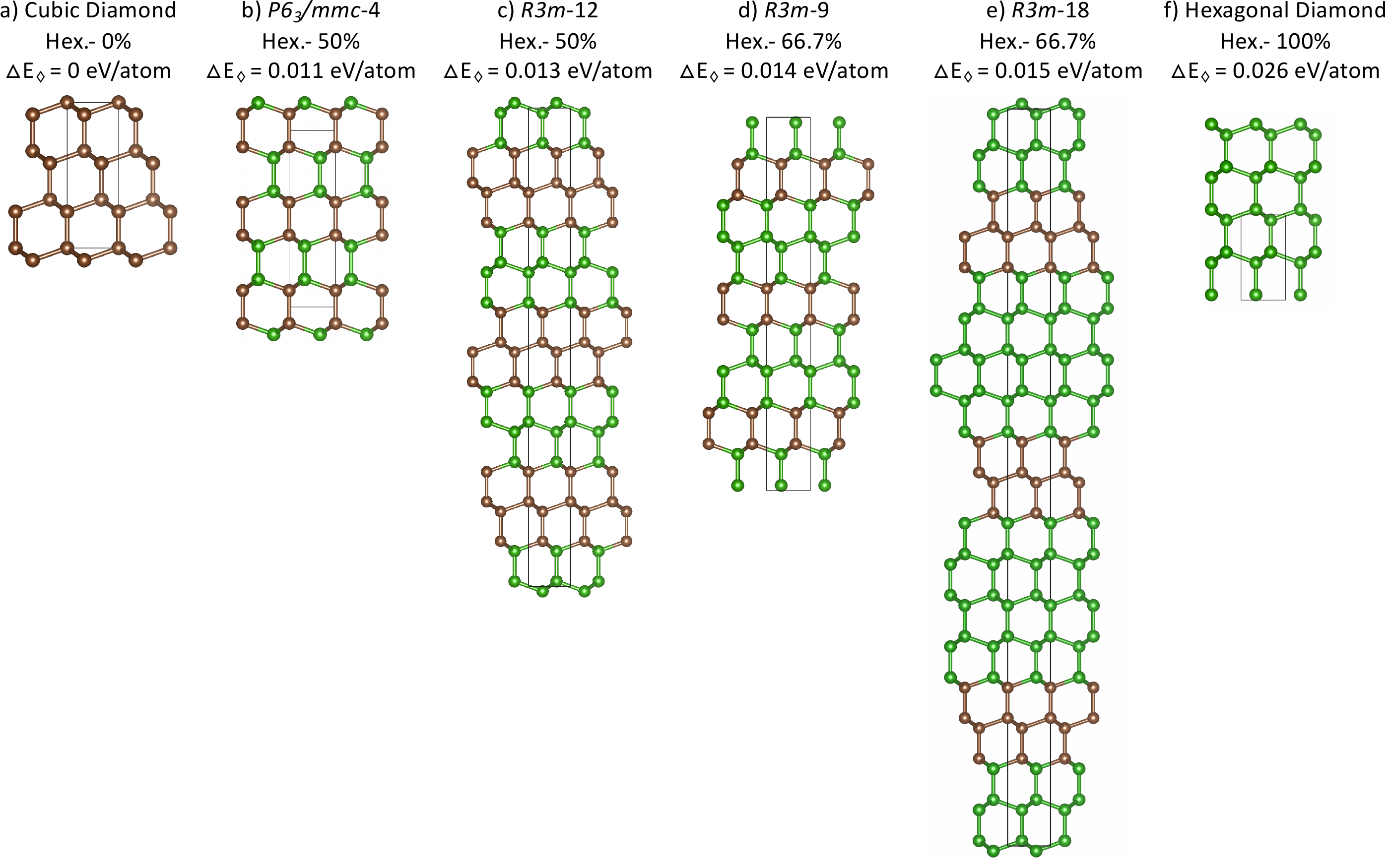}\includegraphics[scale=0.47]{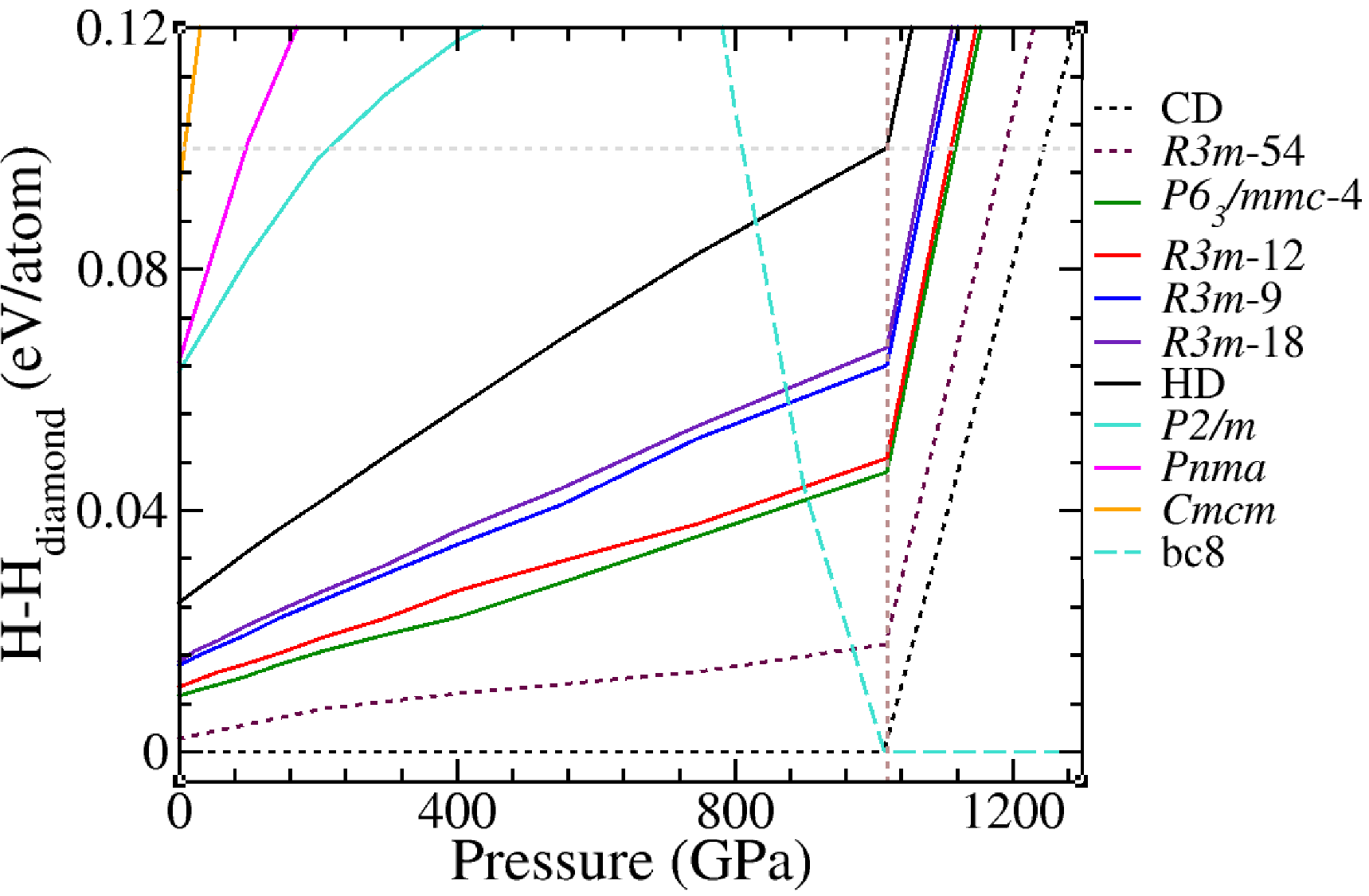}

\caption{\label{fig:USPEX-Results}a) Snapshots of polytypes predicted in first-principles
searches; brown atoms denote CD stacking and green atoms denote HD
stacking. b) Energetics of predicted metastable crystals found across
all searches. The maroon dashed line corresponds to the lowest enthalpy
diamond polytype samples from databases. Reference structure to the
left of the brown vertical dashed line is CD, while BC8 is used as
the reference to the right.}
\end{figure*}

\begin{figure*}
\includegraphics[scale=0.5]{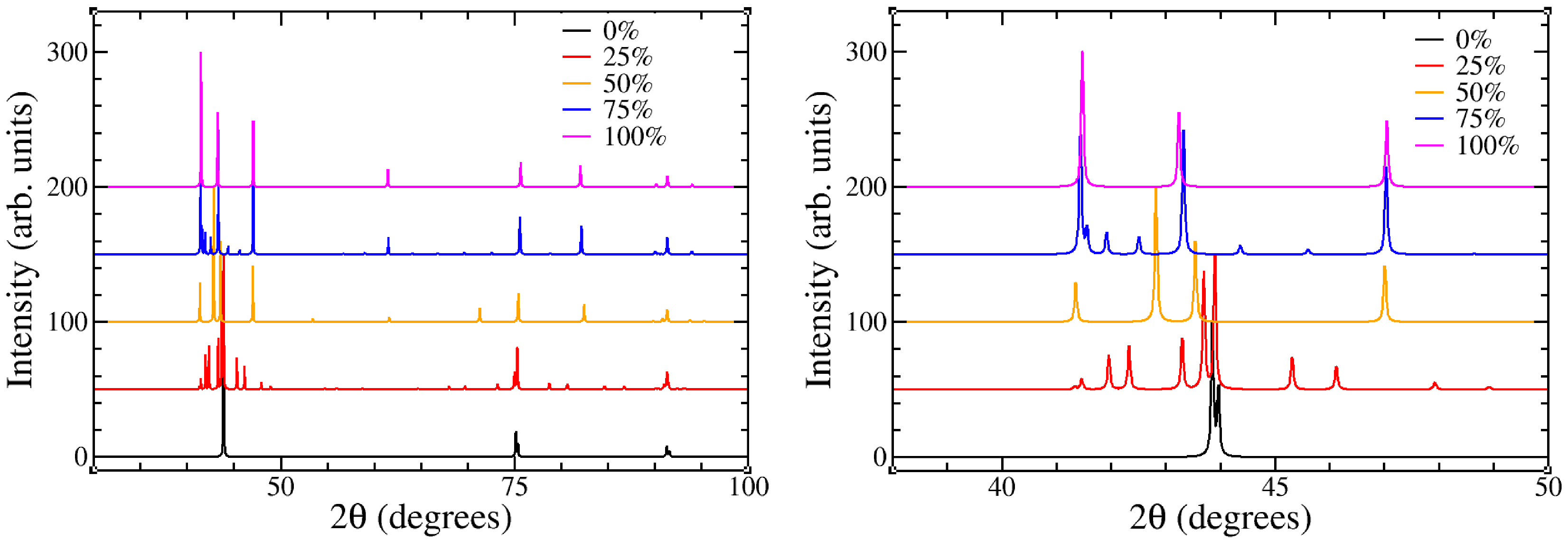}

\caption{\label{fig:XRD}XRD patterns of diamond polytypes.}
\end{figure*}

\begin{figure}
\includegraphics[scale=0.3]{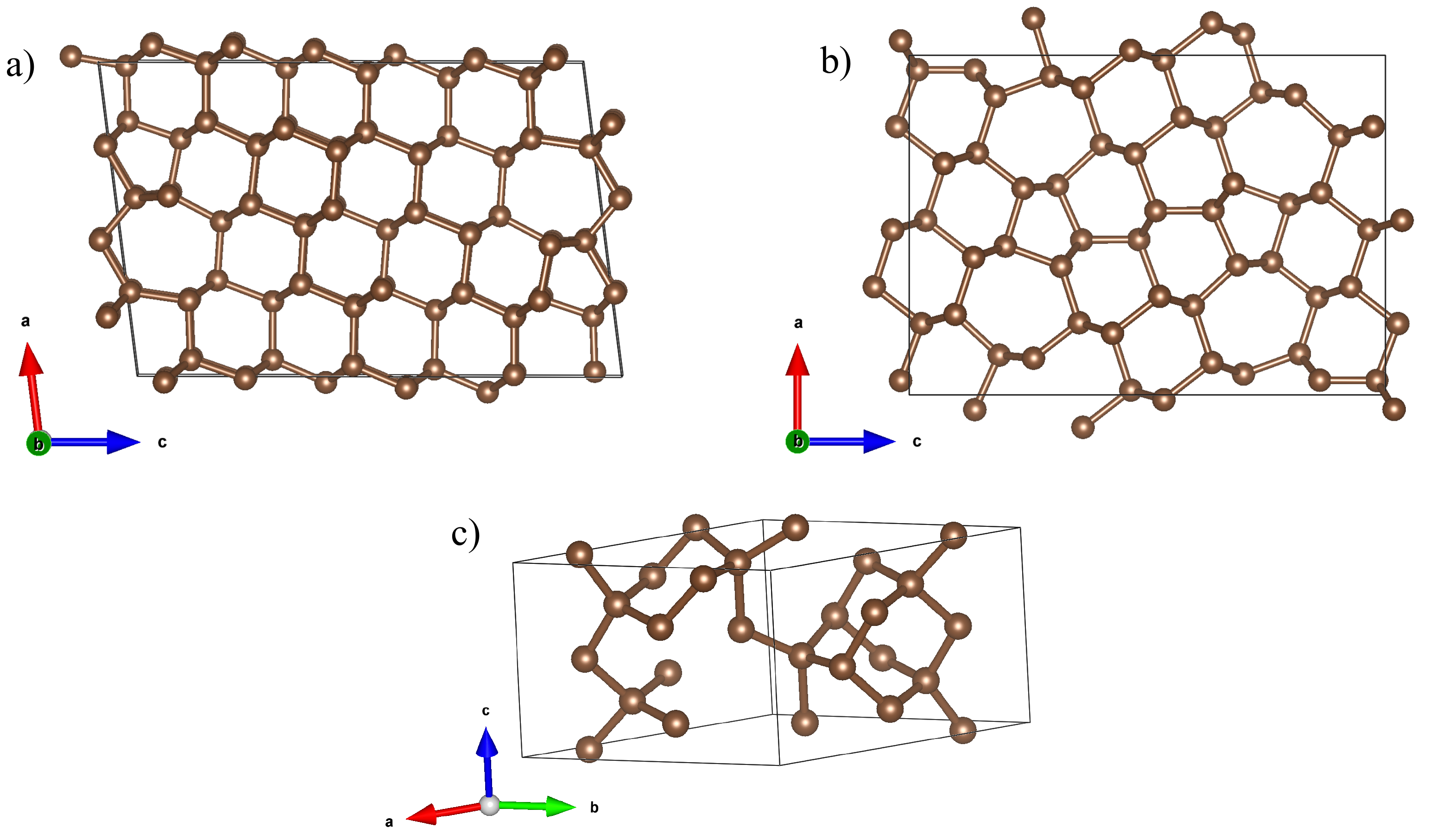}

\caption{\label{fig:3D-Mesh-Snapshots}}
\end{figure}

Four diamond polytypes and HD are predicted across our searches. Two
polytypes, \textit{P$6_{3}$/mmc}-4 and \textit{R3m}-12, have 50\%
hexagonality while the other two polytypes, \textit{R3m}-9 and \textit{R3m}-18,
have 66.7\% hexagonality. Snapshots of the predicted polytypes can
be seen in Fig. \ref{fig:USPEX-Results}a. Beyond the diamond polytypes,
three 3D mesh structures are predicted to be in the metastable region
at low pressures. Snapshots of these crystals are displayed in Figure
S\ref{fig:3D-Mesh-Snapshots}. The enthalpies of these crystals are
much higher than the diamond polytypes so their appearance under shock
compression is unlikely. Mujica \textit{et al.} predicted a crystal
with symmetry \textit{Pbam} that had a similar 5/6/7 membered ring
network to what is seen in Figure S\ref{fig:3D-Mesh-Snapshots}a-b
and also predicted the existence of a new chiral framework with tetragonal
symmetry \textit{P4$_{1}$2$_{1}$2}. We observe this \textit{P4$_{1}$2$_{1}$2}
crystal in our searches as well, however it does not meet our enthalpy
requirement for metastability. 

\begin{figure}
\includegraphics[scale=0.3]{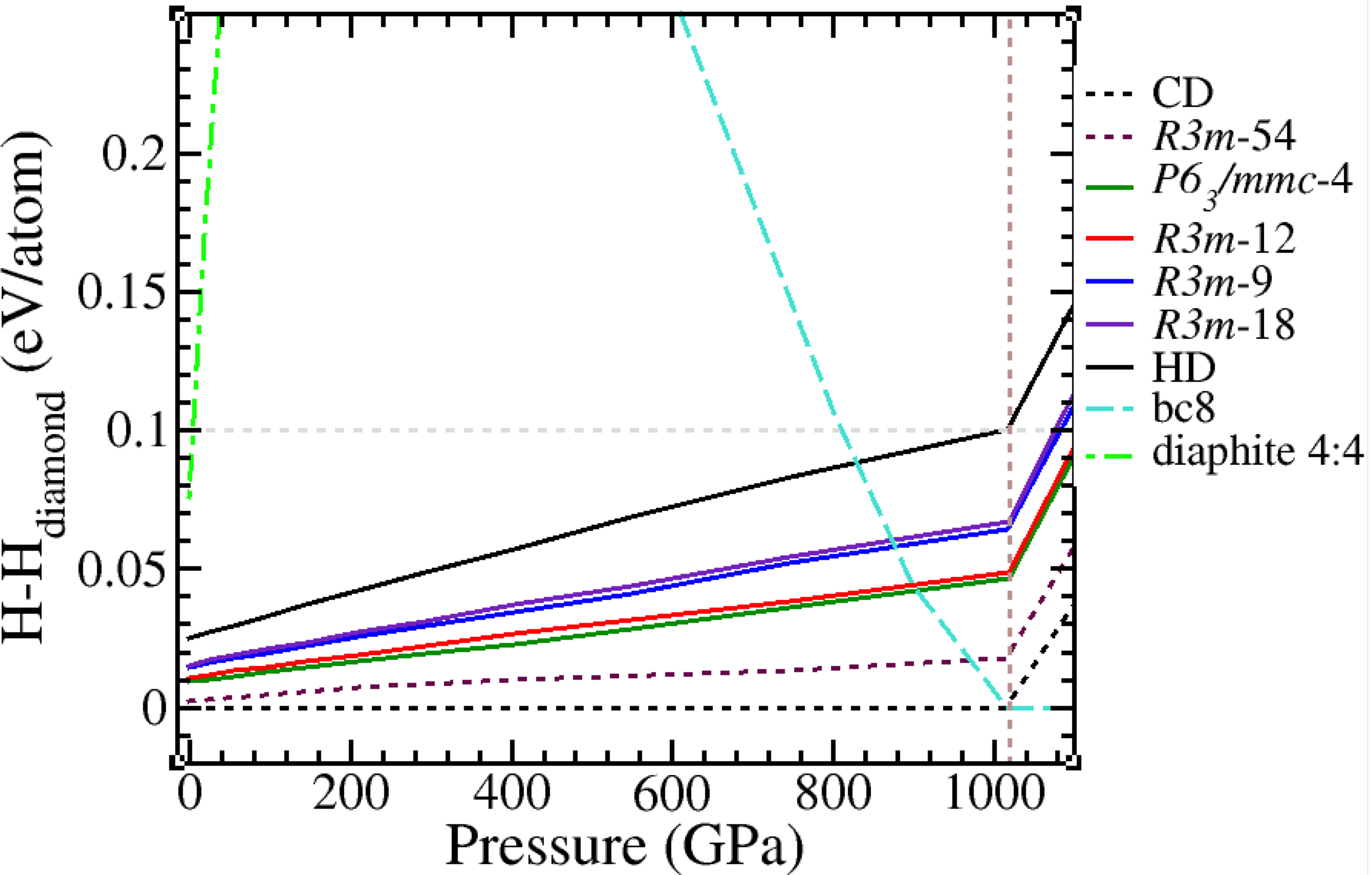}

\caption{\label{fig:Diaphite-Energetics}}
\end{figure}

Past experiments have claimed to observe hexagonal diamond from shock
loading both graphite\citep{Turneaure2017,Stavrou2020,Volz2020,Dong2020,Zhu2020}
and diamond \citep{He2002}, however, there are still questions surrounding
these observations. One major question being posed is whether or not
it is possible to resolve such small differences in the crystals and
definitively say what transformation has occurred. X-ray diffraction
(XRD) patterns for several mixtures of cubic and hexagonal diamond
polytypes (Figure \ref{fig:XRD}) suggest that the differences in
the crystal lattice of various polytypes can indeed be resolved. In
fact, a clear trend can be observed in the XRD as the polytypes increase
in hexagonality. The peak in CD at 44 degrees can be seen to split
as the hexagonality of the crystal grows. In polytypes with hexagonality
greater than 50\%, there are clear peaks that grow at 62 and 82 degrees.
Recently, it has been proposed that diamond-graphene composite structures
will appear under shock wave loading instead of the diamond polytypes
presented here and while HRTEM images do appear to show areas of the
composite structure in impact diamonds confirmation of these findings
via XRD patterns is not conclusive \citep{Nemeth2020,Nemeth2020a}.
Nemeth \textit{et al.} state that the areas of these composite structures
are small\citep{Nemeth2020}, which make capturing them by XRD challenging.
While some composite structures do have a comparable enthalpy difference
to diamond polytypes at very low pressures, the same cannot be said
for these structures at even moderate pressure (see Figure S\ref{fig:Diaphite-Energetics}).
This is consistent with the behavior of graphene, which becomes less
favorable to diamond very rapidly at low pressures. These factors,
combined with the trend we observe in our simulated x-ray diffraction
patterns of diamond polytypes as they grow in hexagonality, lead to
the conclusion that the appearance of the composite structures in
the HRTEM images are the result of anisotropic conditions in the impact
diamond. The class of diamond polytypes detailed in this work should
still be considered the leading candidates for controlled shock experiments
on diamond. 

In conclusion, a unique method of crystal structure prediction has
been employed to explore the effect of shock wave loading on a fixed
unit cell. The previously known hydrostatic phase diagram of carbon
is confirmed via both the new method and the traditional fixed pressure
crystal structure searches. The new method also predicts diamond polytypes
of mixed cubic diamond and hexagonal diamond stacking to be energetically
preferred to conventional uniaxially compressed diamond at longitudinal
stresses exceeding 100 GPa. The existence of these low enthalpy polytypes
suggests that they are the likely candidates for metastable phases
of carbon to appear upon shockwave loading of diamond. It is therefore
unlikely that recently proposed complex crystalline forms \citep{Nemeth2020,Mujica2015,Li2018}
would result from shock experiments. Finally, calculated XRD patterns
show that the resolution of diamond polytypes that result from the
shock compression of diamond should be distinguishable from both cubic
diamond and hexagonal diamond. 

\section*{Methods}

We performed several crystal structure searches using USPEX \citep{Glass2006,Lyakhov2010,Oganov2006},
an evolutionary algorithm that works in generations, to predict the
most energetically favorable crystal for a given environment. The
enthalpy of the crystals in each generation is minimized using the
density functional theory (DFT) code VASP\citep{Kresse:1996} by relaxing
the unit cell parameters and atomic positions for a fixed hydrostatic
pressure. At the end of each generation, the crystals are ranked according
to their enthalpy with the lowest enthalpy individual being the best
candidate for that generation. This cyclic process repeats until the
best individual is kept for ten generations. To investigate the hydrostatic
phase diagram of carbon, searches were performed at fixed pressures
of 100, 500, 1000, 3000, and 5000 GPa. To mimic the environment of
shock-compressed diamond, searches were performed for crystals with
fixed lattice vectors corresponding to 100, 300, 500, and 1000 GPa
longitudinal stress for <100>, <110>, and <111> diamond. During these
searches, only the atoms were relaxed to ensure the uniaxial compression
environment was maintained. See supplemental information for full
methods details.

\bibliographystyle{apsrev4-1}
\bibliography{MetastableCarbon_11-17-2022}

\begin{thebibliography}{35}%
\makeatletter
\providecommand \@ifxundefined [1]{%
 \@ifx{#1\undefined}
}%
\providecommand \@ifnum [1]{%
 \ifnum #1\expandafter \@firstoftwo
 \else \expandafter \@secondoftwo
 \fi
}%
\providecommand \@ifx [1]{%
 \ifx #1\expandafter \@firstoftwo
 \else \expandafter \@secondoftwo
 \fi
}%
\providecommand \natexlab [1]{#1}%
\providecommand \enquote  [1]{``#1''}%
\providecommand \bibnamefont  [1]{#1}%
\providecommand \bibfnamefont [1]{#1}%
\providecommand \citenamefont [1]{#1}%
\providecommand \href@noop [0]{\@secondoftwo}%
\providecommand \href [0]{\begingroup \@sanitize@url \@href}%
\providecommand \@href[1]{\@@startlink{#1}\@@href}%
\providecommand \@@href[1]{\endgroup#1\@@endlink}%
\providecommand \@sanitize@url [0]{\catcode `\\12\catcode `\$12\catcode
  `\&12\catcode `\#12\catcode `\^12\catcode `\_12\catcode `\%12\relax}%
\providecommand \@@startlink[1]{}%
\providecommand \@@endlink[0]{}%
\providecommand \url  [0]{\begingroup\@sanitize@url \@url }%
\providecommand \@url [1]{\endgroup\@href {#1}{\urlprefix }}%
\providecommand \urlprefix  [0]{URL }%
\providecommand \Eprint [0]{\href }%
\providecommand \doibase [0]{http://dx.doi.org/}%
\providecommand \selectlanguage [0]{\@gobble}%
\providecommand \bibinfo  [0]{\@secondoftwo}%
\providecommand \bibfield  [0]{\@secondoftwo}%
\providecommand \translation [1]{[#1]}%
\providecommand \BibitemOpen [0]{}%
\providecommand \bibitemStop [0]{}%
\providecommand \bibitemNoStop [0]{.\EOS\space}%
\providecommand \EOS [0]{\spacefactor3000\relax}%
\providecommand \BibitemShut  [1]{\csname bibitem#1\endcsname}%
\let\auto@bib@innerbib\@empty
%</preamble>
\bibitem [{\citenamefont {Kondo}\ and\ \citenamefont
  {Ahrens}(1983)}]{Kondo1983}%
  \BibitemOpen
  \bibfield  {author} {\bibinfo {author} {\bibfnamefont {K.-I.}\ \bibnamefont
  {Kondo}}\ and\ \bibinfo {author} {\bibfnamefont {T.~J.}\ \bibnamefont
  {Ahrens}},\ }\href {\doibase 10.1029/GL010i004p00281} {\bibfield  {journal}
  {\bibinfo  {journal} {Geophys. Res. Lett.}\ }\textbf {\bibinfo {volume}
  {10}},\ \bibinfo {pages} {281} (\bibinfo {year} {1983})}\BibitemShut
  {NoStop}%
\bibitem [{\citenamefont {Scandolo}\ \emph {et~al.}(1995)\citenamefont
  {Scandolo}, \citenamefont {Bernasconi}, \citenamefont {Chiarotti},
  \citenamefont {Focher},\ and\ \citenamefont {Tosatti}}]{Scandolo1995}%
  \BibitemOpen
  \bibfield  {author} {\bibinfo {author} {\bibfnamefont {S.}~\bibnamefont
  {Scandolo}}, \bibinfo {author} {\bibfnamefont {M.}~\bibnamefont
  {Bernasconi}}, \bibinfo {author} {\bibfnamefont {G.~L.}\ \bibnamefont
  {Chiarotti}}, \bibinfo {author} {\bibfnamefont {P.}~\bibnamefont {Focher}}, \
  and\ \bibinfo {author} {\bibfnamefont {E.}~\bibnamefont {Tosatti}},\ }\href
  {\doibase 10.1103/PhysRevLett.74.4015} {\bibfield  {journal} {\bibinfo
  {journal} {Phys. Rev. Lett.}\ }\textbf {\bibinfo {volume} {74}},\ \bibinfo
  {pages} {4015} (\bibinfo {year} {1995})}\BibitemShut {NoStop}%
\bibitem [{\citenamefont {Knudson}\ \emph {et~al.}(2008)\citenamefont
  {Knudson}, \citenamefont {Desjarlais},\ and\ \citenamefont
  {Dolan}}]{Knudson2008}%
  \BibitemOpen
  \bibfield  {author} {\bibinfo {author} {\bibfnamefont {M.~D.}\ \bibnamefont
  {Knudson}}, \bibinfo {author} {\bibfnamefont {M.~P.}\ \bibnamefont
  {Desjarlais}}, \ and\ \bibinfo {author} {\bibfnamefont {D.~H.}\ \bibnamefont
  {Dolan}},\ }\href {\doibase 10.1126/science.1165278} {\bibfield  {journal}
  {\bibinfo  {journal} {Science}\ }\textbf {\bibinfo {volume} {322}},\ \bibinfo
  {pages} {1822} (\bibinfo {year} {2008})}\BibitemShut {NoStop}%
\bibitem [{\citenamefont {Eggert}\ \emph {et~al.}(2010)\citenamefont {Eggert},
  \citenamefont {Hicks}, \citenamefont {Celliers}, \citenamefont {Bradley},
  \citenamefont {McWilliams}, \citenamefont {Jeanloz}, \citenamefont {Miller},
  \citenamefont {Boehly},\ and\ \citenamefont {Collins}}]{Eggert2009}%
  \BibitemOpen
  \bibfield  {author} {\bibinfo {author} {\bibfnamefont {J.~H.}\ \bibnamefont
  {Eggert}}, \bibinfo {author} {\bibfnamefont {D.~G.}\ \bibnamefont {Hicks}},
  \bibinfo {author} {\bibfnamefont {P.~M.}\ \bibnamefont {Celliers}}, \bibinfo
  {author} {\bibfnamefont {D.~K.}\ \bibnamefont {Bradley}}, \bibinfo {author}
  {\bibfnamefont {R.~S.}\ \bibnamefont {McWilliams}}, \bibinfo {author}
  {\bibfnamefont {R.}~\bibnamefont {Jeanloz}}, \bibinfo {author} {\bibfnamefont
  {J.~E.}\ \bibnamefont {Miller}}, \bibinfo {author} {\bibfnamefont {T.~R.}\
  \bibnamefont {Boehly}}, \ and\ \bibinfo {author} {\bibfnamefont {G.~W.}\
  \bibnamefont {Collins}},\ }\href {\doibase 10.1038/nphys1438} {\bibfield
  {journal} {\bibinfo  {journal} {Nat. Phys.}\ }\textbf {\bibinfo {volume}
  {6}},\ \bibinfo {pages} {40} (\bibinfo {year} {2010})}\BibitemShut {NoStop}%
\bibitem [{\citenamefont {McWilliams}\ \emph {et~al.}(2010)\citenamefont
  {McWilliams}, \citenamefont {Eggert}, \citenamefont {Hicks}, \citenamefont
  {Bradley}, \citenamefont {Celliers}, \citenamefont {Spaulding}, \citenamefont
  {Boehly}, \citenamefont {Collins},\ and\ \citenamefont
  {Jeanloz}}]{McWilliams2010}%
  \BibitemOpen
  \bibfield  {author} {\bibinfo {author} {\bibfnamefont {R.~S.}\ \bibnamefont
  {McWilliams}}, \bibinfo {author} {\bibfnamefont {J.~H.}\ \bibnamefont
  {Eggert}}, \bibinfo {author} {\bibfnamefont {D.~G.}\ \bibnamefont {Hicks}},
  \bibinfo {author} {\bibfnamefont {D.~K.}\ \bibnamefont {Bradley}}, \bibinfo
  {author} {\bibfnamefont {P.~M.}\ \bibnamefont {Celliers}}, \bibinfo {author}
  {\bibfnamefont {D.~K.}\ \bibnamefont {Spaulding}}, \bibinfo {author}
  {\bibfnamefont {T.~R.}\ \bibnamefont {Boehly}}, \bibinfo {author}
  {\bibfnamefont {G.~W.}\ \bibnamefont {Collins}}, \ and\ \bibinfo {author}
  {\bibfnamefont {R.}~\bibnamefont {Jeanloz}},\ }\href {\doibase
  10.1103/PhysRevB.81.014111} {\bibfield  {journal} {\bibinfo  {journal} {Phys.
  Rev. B}\ }\textbf {\bibinfo {volume} {81}},\ \bibinfo {pages} {014111}
  (\bibinfo {year} {2010})}\BibitemShut {NoStop}%
\bibitem [{\citenamefont {Kraus}\ \emph {et~al.}(2012)\citenamefont {Kraus},
  \citenamefont {Otten}, \citenamefont {Frank}, \citenamefont {Bagnoud},
  \citenamefont {Bla{\v{z}}evi{\'{c}}}, \citenamefont {Gericke}, \citenamefont
  {Gregori}, \citenamefont {Ortner}, \citenamefont {Schaumann}, \citenamefont
  {Schumacher}, \citenamefont {Vorberger}, \citenamefont {Wagner},
  \citenamefont {W{\"{u}}nsch},\ and\ \citenamefont {Roth}}]{Kraus2012}%
  \BibitemOpen
  \bibfield  {author} {\bibinfo {author} {\bibfnamefont {D.}~\bibnamefont
  {Kraus}}, \bibinfo {author} {\bibfnamefont {A.}~\bibnamefont {Otten}},
  \bibinfo {author} {\bibfnamefont {A.}~\bibnamefont {Frank}}, \bibinfo
  {author} {\bibfnamefont {V.}~\bibnamefont {Bagnoud}}, \bibinfo {author}
  {\bibfnamefont {A.}~\bibnamefont {Bla{\v{z}}evi{\'{c}}}}, \bibinfo {author}
  {\bibfnamefont {D.}~\bibnamefont {Gericke}}, \bibinfo {author} {\bibfnamefont
  {G.}~\bibnamefont {Gregori}}, \bibinfo {author} {\bibfnamefont
  {A.}~\bibnamefont {Ortner}}, \bibinfo {author} {\bibfnamefont
  {G.}~\bibnamefont {Schaumann}}, \bibinfo {author} {\bibfnamefont
  {D.}~\bibnamefont {Schumacher}}, \bibinfo {author} {\bibfnamefont
  {J.}~\bibnamefont {Vorberger}}, \bibinfo {author} {\bibfnamefont
  {F.}~\bibnamefont {Wagner}}, \bibinfo {author} {\bibfnamefont
  {K.}~\bibnamefont {W{\"{u}}nsch}}, \ and\ \bibinfo {author} {\bibfnamefont
  {M.}~\bibnamefont {Roth}},\ }\href {\doibase 10.1016/j.hedp.2011.11.011}
  {\bibfield  {journal} {\bibinfo  {journal} {High Energy Density Phys.}\
  }\textbf {\bibinfo {volume} {8}},\ \bibinfo {pages} {46} (\bibinfo {year}
  {2012})}\BibitemShut {NoStop}%
\bibitem [{\citenamefont {Smith}\ \emph {et~al.}(2014)\citenamefont {Smith},
  \citenamefont {Eggert}, \citenamefont {Jeanloz}, \citenamefont {Duffy},
  \citenamefont {Braun}, \citenamefont {Patterson}, \citenamefont {Rudd},
  \citenamefont {Biener}, \citenamefont {Lazicki}, \citenamefont {Hamza},
  \citenamefont {Wang}, \citenamefont {Braun}, \citenamefont {Benedict},
  \citenamefont {Celliers},\ and\ \citenamefont {Collins}}]{Smith2014}%
  \BibitemOpen
  \bibfield  {author} {\bibinfo {author} {\bibfnamefont {R.~F.}\ \bibnamefont
  {Smith}}, \bibinfo {author} {\bibfnamefont {J.~H.}\ \bibnamefont {Eggert}},
  \bibinfo {author} {\bibfnamefont {R.}~\bibnamefont {Jeanloz}}, \bibinfo
  {author} {\bibfnamefont {T.~S.}\ \bibnamefont {Duffy}}, \bibinfo {author}
  {\bibfnamefont {D.~G.}\ \bibnamefont {Braun}}, \bibinfo {author}
  {\bibfnamefont {J.~R.}\ \bibnamefont {Patterson}}, \bibinfo {author}
  {\bibfnamefont {R.~E.}\ \bibnamefont {Rudd}}, \bibinfo {author}
  {\bibfnamefont {J.}~\bibnamefont {Biener}}, \bibinfo {author} {\bibfnamefont
  {A.~E.}\ \bibnamefont {Lazicki}}, \bibinfo {author} {\bibfnamefont {A.~V.}\
  \bibnamefont {Hamza}}, \bibinfo {author} {\bibfnamefont {J.}~\bibnamefont
  {Wang}}, \bibinfo {author} {\bibfnamefont {T.}~\bibnamefont {Braun}},
  \bibinfo {author} {\bibfnamefont {L.~X.}\ \bibnamefont {Benedict}}, \bibinfo
  {author} {\bibfnamefont {P.~M.}\ \bibnamefont {Celliers}}, \ and\ \bibinfo
  {author} {\bibfnamefont {G.~W.}\ \bibnamefont {Collins}},\ }\href {\doibase
  10.1038/nature13526} {\bibfield  {journal} {\bibinfo  {journal} {Nature}\
  }\textbf {\bibinfo {volume} {511}},\ \bibinfo {pages} {330} (\bibinfo {year}
  {2014})}\BibitemShut {NoStop}%
\bibitem [{\citenamefont {Jones}\ \emph {et~al.}(2016)\citenamefont {Jones},
  \citenamefont {McMillan}, \citenamefont {Salzmann}, \citenamefont {Alvaro},
  \citenamefont {Nestola}, \citenamefont {Prencipe}, \citenamefont {Dobson},
  \citenamefont {Hazael},\ and\ \citenamefont {Moore}}]{Jones2016}%
  \BibitemOpen
  \bibfield  {author} {\bibinfo {author} {\bibfnamefont {A.~P.}\ \bibnamefont
  {Jones}}, \bibinfo {author} {\bibfnamefont {P.~F.}\ \bibnamefont {McMillan}},
  \bibinfo {author} {\bibfnamefont {C.~G.}\ \bibnamefont {Salzmann}}, \bibinfo
  {author} {\bibfnamefont {M.}~\bibnamefont {Alvaro}}, \bibinfo {author}
  {\bibfnamefont {F.}~\bibnamefont {Nestola}}, \bibinfo {author} {\bibfnamefont
  {M.}~\bibnamefont {Prencipe}}, \bibinfo {author} {\bibfnamefont
  {D.}~\bibnamefont {Dobson}}, \bibinfo {author} {\bibfnamefont
  {R.}~\bibnamefont {Hazael}}, \ and\ \bibinfo {author} {\bibfnamefont
  {M.}~\bibnamefont {Moore}},\ }\href {\doibase 10.1016/j.lithos.2016.09.023}
  {\bibfield  {journal} {\bibinfo  {journal} {Lithos}\ }\textbf {\bibinfo
  {volume} {265}},\ \bibinfo {pages} {214} (\bibinfo {year}
  {2016})}\BibitemShut {NoStop}%
\bibitem [{\citenamefont {N{\'{e}}meth}\ \emph
  {et~al.}(2020{\natexlab{a}})\citenamefont {N{\'{e}}meth}, \citenamefont
  {McColl}, \citenamefont {Smith}, \citenamefont {Murri}, \citenamefont
  {Garvie}, \citenamefont {Alvaro}, \citenamefont {P{\'{e}}cz}, \citenamefont
  {Jones}, \citenamefont {Cor{\`{a}}}, \citenamefont {Salzmann},\ and\
  \citenamefont {McMillan}}]{Nemeth2020}%
  \BibitemOpen
  \bibfield  {author} {\bibinfo {author} {\bibfnamefont {P.}~\bibnamefont
  {N{\'{e}}meth}}, \bibinfo {author} {\bibfnamefont {K.}~\bibnamefont
  {McColl}}, \bibinfo {author} {\bibfnamefont {R.~L.}\ \bibnamefont {Smith}},
  \bibinfo {author} {\bibfnamefont {M.}~\bibnamefont {Murri}}, \bibinfo
  {author} {\bibfnamefont {L.~A.~J.}\ \bibnamefont {Garvie}}, \bibinfo {author}
  {\bibfnamefont {M.}~\bibnamefont {Alvaro}}, \bibinfo {author} {\bibfnamefont
  {B.}~\bibnamefont {P{\'{e}}cz}}, \bibinfo {author} {\bibfnamefont {A.~P.}\
  \bibnamefont {Jones}}, \bibinfo {author} {\bibfnamefont {F.}~\bibnamefont
  {Cor{\`{a}}}}, \bibinfo {author} {\bibfnamefont {C.~G.}\ \bibnamefont
  {Salzmann}}, \ and\ \bibinfo {author} {\bibfnamefont {P.~F.}\ \bibnamefont
  {McMillan}},\ }\href {\doibase 10.1021/acs.nanolett.0c00556} {\bibfield
  {journal} {\bibinfo  {journal} {Nano Lett.}\ }\textbf {\bibinfo {volume}
  {20}},\ \bibinfo {pages} {3611} (\bibinfo {year}
  {2020}{\natexlab{a}})}\BibitemShut {NoStop}%
\bibitem [{\citenamefont {N{\'{e}}meth}\ \emph
  {et~al.}(2020{\natexlab{b}})\citenamefont {N{\'{e}}meth}, \citenamefont
  {McColl}, \citenamefont {Garvie}, \citenamefont {Salzmann}, \citenamefont
  {Murri},\ and\ \citenamefont {McMillan}}]{Nemeth2020a}%
  \BibitemOpen
  \bibfield  {author} {\bibinfo {author} {\bibfnamefont {P.}~\bibnamefont
  {N{\'{e}}meth}}, \bibinfo {author} {\bibfnamefont {K.}~\bibnamefont
  {McColl}}, \bibinfo {author} {\bibfnamefont {L.~A.~J.}\ \bibnamefont
  {Garvie}}, \bibinfo {author} {\bibfnamefont {C.~G.}\ \bibnamefont
  {Salzmann}}, \bibinfo {author} {\bibfnamefont {M.}~\bibnamefont {Murri}}, \
  and\ \bibinfo {author} {\bibfnamefont {P.~F.}\ \bibnamefont {McMillan}},\
  }\href {\doibase 10.1038/s41563-020-0759-8} {\bibfield  {journal} {\bibinfo
  {journal} {Nat. Mater.}\ }\textbf {\bibinfo {volume} {19}},\ \bibinfo {pages}
  {1126} (\bibinfo {year} {2020}{\natexlab{b}})}\BibitemShut {NoStop}%
\bibitem [{\citenamefont {Turneaure}\ \emph {et~al.}(2017)\citenamefont
  {Turneaure}, \citenamefont {Sharma}, \citenamefont {Volz}, \citenamefont
  {Winey},\ and\ \citenamefont {Gupta}}]{Turneaure2017}%
  \BibitemOpen
  \bibfield  {author} {\bibinfo {author} {\bibfnamefont {S.~J.}\ \bibnamefont
  {Turneaure}}, \bibinfo {author} {\bibfnamefont {S.~M.}\ \bibnamefont
  {Sharma}}, \bibinfo {author} {\bibfnamefont {T.~J.}\ \bibnamefont {Volz}},
  \bibinfo {author} {\bibfnamefont {J.~M.}\ \bibnamefont {Winey}}, \ and\
  \bibinfo {author} {\bibfnamefont {Y.~M.}\ \bibnamefont {Gupta}},\ }\href
  {\doibase 10.1126/sciadv.aao3561} {\bibfield  {journal} {\bibinfo  {journal}
  {Sci. Adv.}\ }\textbf {\bibinfo {volume} {3}},\ \bibinfo {pages} {eaao3561}
  (\bibinfo {year} {2017})}\BibitemShut {NoStop}%
\bibitem [{\citenamefont {Stavrou}\ \emph {et~al.}(2020)\citenamefont
  {Stavrou}, \citenamefont {Bagge-Hansen}, \citenamefont {Hammons},
  \citenamefont {Nielsen}, \citenamefont {Steele}, \citenamefont {Xiao},
  \citenamefont {Kroonblawd}, \citenamefont {Nelms}, \citenamefont {Shaw},
  \citenamefont {Bassett}, \citenamefont {Bastea}, \citenamefont {Lauderbach},
  \citenamefont {Hodgin}, \citenamefont {Perez-Marty}, \citenamefont {Singh},
  \citenamefont {Das}, \citenamefont {Li}, \citenamefont {Schuman},
  \citenamefont {Sinclair}, \citenamefont {Fezzaa}, \citenamefont {Deriy},
  \citenamefont {Leininger},\ and\ \citenamefont {Willey}}]{Stavrou2020}%
  \BibitemOpen
  \bibfield  {author} {\bibinfo {author} {\bibfnamefont {E.}~\bibnamefont
  {Stavrou}}, \bibinfo {author} {\bibfnamefont {M.}~\bibnamefont
  {Bagge-Hansen}}, \bibinfo {author} {\bibfnamefont {J.~A.}\ \bibnamefont
  {Hammons}}, \bibinfo {author} {\bibfnamefont {M.~H.}\ \bibnamefont
  {Nielsen}}, \bibinfo {author} {\bibfnamefont {B.~A.}\ \bibnamefont {Steele}},
  \bibinfo {author} {\bibfnamefont {P.}~\bibnamefont {Xiao}}, \bibinfo {author}
  {\bibfnamefont {M.~P.}\ \bibnamefont {Kroonblawd}}, \bibinfo {author}
  {\bibfnamefont {M.~D.}\ \bibnamefont {Nelms}}, \bibinfo {author}
  {\bibfnamefont {W.~L.}\ \bibnamefont {Shaw}}, \bibinfo {author}
  {\bibfnamefont {W.}~\bibnamefont {Bassett}}, \bibinfo {author} {\bibfnamefont
  {S.}~\bibnamefont {Bastea}}, \bibinfo {author} {\bibfnamefont {L.~M.}\
  \bibnamefont {Lauderbach}}, \bibinfo {author} {\bibfnamefont {R.~L.}\
  \bibnamefont {Hodgin}}, \bibinfo {author} {\bibfnamefont {N.~A.}\
  \bibnamefont {Perez-Marty}}, \bibinfo {author} {\bibfnamefont
  {S.}~\bibnamefont {Singh}}, \bibinfo {author} {\bibfnamefont
  {P.}~\bibnamefont {Das}}, \bibinfo {author} {\bibfnamefont {Y.}~\bibnamefont
  {Li}}, \bibinfo {author} {\bibfnamefont {A.}~\bibnamefont {Schuman}},
  \bibinfo {author} {\bibfnamefont {N.}~\bibnamefont {Sinclair}}, \bibinfo
  {author} {\bibfnamefont {K.}~\bibnamefont {Fezzaa}}, \bibinfo {author}
  {\bibfnamefont {A.}~\bibnamefont {Deriy}}, \bibinfo {author} {\bibfnamefont
  {L.~D.}\ \bibnamefont {Leininger}}, \ and\ \bibinfo {author} {\bibfnamefont
  {T.~M.}\ \bibnamefont {Willey}},\ }\href {\doibase
  10.1103/PhysRevB.102.104116} {\bibfield  {journal} {\bibinfo  {journal}
  {Phys. Rev. B}\ }\textbf {\bibinfo {volume} {102}},\ \bibinfo {pages}
  {104116} (\bibinfo {year} {2020})}\BibitemShut {NoStop}%
\bibitem [{\citenamefont {Volz}\ \emph {et~al.}(2020)\citenamefont {Volz},
  \citenamefont {Turneaure}, \citenamefont {Sharma},\ and\ \citenamefont
  {Gupta}}]{Volz2020}%
  \BibitemOpen
  \bibfield  {author} {\bibinfo {author} {\bibfnamefont {T.~J.}\ \bibnamefont
  {Volz}}, \bibinfo {author} {\bibfnamefont {S.~J.}\ \bibnamefont {Turneaure}},
  \bibinfo {author} {\bibfnamefont {S.~M.}\ \bibnamefont {Sharma}}, \ and\
  \bibinfo {author} {\bibfnamefont {Y.~M.}\ \bibnamefont {Gupta}},\ }\href
  {\doibase 10.1103/PhysRevB.101.224109} {\bibfield  {journal} {\bibinfo
  {journal} {Phys. Rev. B}\ }\textbf {\bibinfo {volume} {101}},\ \bibinfo
  {pages} {224109} (\bibinfo {year} {2020})}\BibitemShut {NoStop}%
\bibitem [{\citenamefont {Correa}\ \emph {et~al.}(2008)\citenamefont {Correa},
  \citenamefont {Benedict}, \citenamefont {Young}, \citenamefont {Schwegler},\
  and\ \citenamefont {Bonev}}]{Correa2008}%
  \BibitemOpen
  \bibfield  {author} {\bibinfo {author} {\bibfnamefont {A.~A.}\ \bibnamefont
  {Correa}}, \bibinfo {author} {\bibfnamefont {L.~X.}\ \bibnamefont
  {Benedict}}, \bibinfo {author} {\bibfnamefont {D.~A.}\ \bibnamefont {Young}},
  \bibinfo {author} {\bibfnamefont {E.}~\bibnamefont {Schwegler}}, \ and\
  \bibinfo {author} {\bibfnamefont {S.~A.}\ \bibnamefont {Bonev}},\ }\href
  {\doibase 10.1103/PhysRevB.78.024101} {\bibfield  {journal} {\bibinfo
  {journal} {Phys. Rev. B}\ }\textbf {\bibinfo {volume} {78}},\ \bibinfo
  {pages} {024101} (\bibinfo {year} {2008})}\BibitemShut {NoStop}%
\bibitem [{\citenamefont {Oleynik}\ \emph {et~al.}(2008)\citenamefont
  {Oleynik}, \citenamefont {Landerville}, \citenamefont {Zybin}, \citenamefont
  {Elert},\ and\ \citenamefont {White}}]{Oleynik2008}%
  \BibitemOpen
  \bibfield  {author} {\bibinfo {author} {\bibfnamefont {I.~I.}\ \bibnamefont
  {Oleynik}}, \bibinfo {author} {\bibfnamefont {A.~C.}\ \bibnamefont
  {Landerville}}, \bibinfo {author} {\bibfnamefont {S.~V.}\ \bibnamefont
  {Zybin}}, \bibinfo {author} {\bibfnamefont {M.~L.}\ \bibnamefont {Elert}}, \
  and\ \bibinfo {author} {\bibfnamefont {C.~T.}\ \bibnamefont {White}},\ }\href
  {\doibase 10.1103/PhysRevB.78.180101} {\bibfield  {journal} {\bibinfo
  {journal} {Phys. Rev. B}\ }\textbf {\bibinfo {volume} {78}},\ \bibinfo
  {pages} {180101} (\bibinfo {year} {2008})}\BibitemShut {NoStop}%
\bibitem [{\citenamefont {Wen}\ \emph {et~al.}(2008)\citenamefont {Wen},
  \citenamefont {Zhao}, \citenamefont {Bucknum}, \citenamefont {Yao},\ and\
  \citenamefont {Li}}]{Wen2008}%
  \BibitemOpen
  \bibfield  {author} {\bibinfo {author} {\bibfnamefont {B.}~\bibnamefont
  {Wen}}, \bibinfo {author} {\bibfnamefont {J.}~\bibnamefont {Zhao}}, \bibinfo
  {author} {\bibfnamefont {M.~J.}\ \bibnamefont {Bucknum}}, \bibinfo {author}
  {\bibfnamefont {P.}~\bibnamefont {Yao}}, \ and\ \bibinfo {author}
  {\bibfnamefont {T.}~\bibnamefont {Li}},\ }\href {\doibase
  10.1016/j.diamond.2008.01.020} {\bibfield  {journal} {\bibinfo  {journal}
  {Diam. Relat. Mater.}\ }\textbf {\bibinfo {volume} {17}},\ \bibinfo {pages}
  {356} (\bibinfo {year} {2008})}\BibitemShut {NoStop}%
\bibitem [{\citenamefont {Sun}\ \emph {et~al.}(2009)\citenamefont {Sun},
  \citenamefont {Klug},\ and\ \citenamefont {Marto?ák}}]{Sun2009}%
  \BibitemOpen
  \bibfield  {author} {\bibinfo {author} {\bibfnamefont {J.}~\bibnamefont
  {Sun}}, \bibinfo {author} {\bibfnamefont {D.~D.}\ \bibnamefont {Klug}}, \
  and\ \bibinfo {author} {\bibfnamefont {R.}~\bibnamefont {Marto?ák}},\ }\href
  {\doibase 10.1063/1.3139060} {\bibfield  {journal} {\bibinfo  {journal} {J.
  Chem. Phys.}\ }\textbf {\bibinfo {volume} {130}},\ \bibinfo {pages} {194512}
  (\bibinfo {year} {2009})}\BibitemShut {NoStop}%
\bibitem [{\citenamefont {Zhu}\ \emph {et~al.}(2011)\citenamefont {Zhu},
  \citenamefont {Oganov}, \citenamefont {Salvad{\'{o}}}, \citenamefont
  {Pertierra},\ and\ \citenamefont {Lyakhov}}]{Zhu2011}%
  \BibitemOpen
  \bibfield  {author} {\bibinfo {author} {\bibfnamefont {Q.}~\bibnamefont
  {Zhu}}, \bibinfo {author} {\bibfnamefont {A.~R.}\ \bibnamefont {Oganov}},
  \bibinfo {author} {\bibfnamefont {M.~A.}\ \bibnamefont {Salvad{\'{o}}}},
  \bibinfo {author} {\bibfnamefont {P.}~\bibnamefont {Pertierra}}, \ and\
  \bibinfo {author} {\bibfnamefont {A.~O.}\ \bibnamefont {Lyakhov}},\ }\href
  {\doibase 10.1103/PhysRevB.83.193410} {\bibfield  {journal} {\bibinfo
  {journal} {Phys. Rev. B}\ }\textbf {\bibinfo {volume} {83}},\ \bibinfo
  {pages} {193410} (\bibinfo {year} {2011})}\BibitemShut {NoStop}%
\bibitem [{\citenamefont {Martinez-Canales}\ \emph {et~al.}(2012)\citenamefont
  {Martinez-Canales}, \citenamefont {Pickard},\ and\ \citenamefont
  {Needs}}]{Martinez-Canales2012}%
  \BibitemOpen
  \bibfield  {author} {\bibinfo {author} {\bibfnamefont {M.}~\bibnamefont
  {Martinez-Canales}}, \bibinfo {author} {\bibfnamefont {C.~J.}\ \bibnamefont
  {Pickard}}, \ and\ \bibinfo {author} {\bibfnamefont {R.~J.}\ \bibnamefont
  {Needs}},\ }\href {\doibase 10.1103/PhysRevLett.108.045704} {\bibfield
  {journal} {\bibinfo  {journal} {Phys. Rev. Lett.}\ }\textbf {\bibinfo
  {volume} {108}},\ \bibinfo {pages} {045704} (\bibinfo {year}
  {2012})}\BibitemShut {NoStop}%
\bibitem [{\citenamefont {Oganov}\ \emph {et~al.}(2013)\citenamefont {Oganov},
  \citenamefont {Hemley}, \citenamefont {Hazen},\ and\ \citenamefont
  {Jones}}]{Oganov2013}%
  \BibitemOpen
  \bibfield  {author} {\bibinfo {author} {\bibfnamefont {A.~R.}\ \bibnamefont
  {Oganov}}, \bibinfo {author} {\bibfnamefont {R.~J.}\ \bibnamefont {Hemley}},
  \bibinfo {author} {\bibfnamefont {R.~M.}\ \bibnamefont {Hazen}}, \ and\
  \bibinfo {author} {\bibfnamefont {A.~P.}\ \bibnamefont {Jones}},\ }\href
  {\doibase 10.2138/rmg.2013.75.3} {\bibfield  {journal} {\bibinfo  {journal}
  {Rev. Mineral. Geochemistry}\ }\textbf {\bibinfo {volume} {75}},\ \bibinfo
  {pages} {47} (\bibinfo {year} {2013})}\BibitemShut {NoStop}%
\bibitem [{\citenamefont {Pineau}(2013)}]{Pineau2013}%
  \BibitemOpen
  \bibfield  {author} {\bibinfo {author} {\bibfnamefont {N.}~\bibnamefont
  {Pineau}},\ }\href {\doibase 10.1021/jp403568m} {\bibfield  {journal}
  {\bibinfo  {journal} {J. Phys. Chem. C}\ }\textbf {\bibinfo {volume} {117}},\
  \bibinfo {pages} {12778} (\bibinfo {year} {2013})}\BibitemShut {NoStop}%
\bibitem [{\citenamefont {Cui}\ \emph {et~al.}(2015)\citenamefont {Cui},
  \citenamefont {Sheng}, \citenamefont {Yan}, \citenamefont {Zhu},
  \citenamefont {Zheng},\ and\ \citenamefont {Su}}]{Cui2015}%
  \BibitemOpen
  \bibfield  {author} {\bibinfo {author} {\bibfnamefont {H.-J.}\ \bibnamefont
  {Cui}}, \bibinfo {author} {\bibfnamefont {X.-L.}\ \bibnamefont {Sheng}},
  \bibinfo {author} {\bibfnamefont {Q.-B.}\ \bibnamefont {Yan}}, \bibinfo
  {author} {\bibfnamefont {Z.-G.}\ \bibnamefont {Zhu}}, \bibinfo {author}
  {\bibfnamefont {Q.-R.}\ \bibnamefont {Zheng}}, \ and\ \bibinfo {author}
  {\bibfnamefont {G.}~\bibnamefont {Su}},\ }\href {\doibase
  10.1016/j.commatsci.2014.09.050} {\bibfield  {journal} {\bibinfo  {journal}
  {Comput. Mater. Sci.}\ }\textbf {\bibinfo {volume} {98}},\ \bibinfo {pages}
  {129} (\bibinfo {year} {2015})}\BibitemShut {NoStop}%
\bibitem [{\citenamefont {Mujica}\ \emph {et~al.}(2015)\citenamefont {Mujica},
  \citenamefont {Pickard},\ and\ \citenamefont {Needs}}]{Mujica2015}%
  \BibitemOpen
  \bibfield  {author} {\bibinfo {author} {\bibfnamefont {A.}~\bibnamefont
  {Mujica}}, \bibinfo {author} {\bibfnamefont {C.~J.}\ \bibnamefont {Pickard}},
  \ and\ \bibinfo {author} {\bibfnamefont {R.~J.}\ \bibnamefont {Needs}},\
  }\href {\doibase 10.1103/PhysRevB.91.214104} {\bibfield  {journal} {\bibinfo
  {journal} {Phys. Rev. B}\ }\textbf {\bibinfo {volume} {91}},\ \bibinfo
  {pages} {214104} (\bibinfo {year} {2015})},\ \Eprint
  {http://arxiv.org/abs/1508.02631} {arXiv:1508.02631} \BibitemShut {NoStop}%
\bibitem [{\citenamefont {Li}\ \emph {et~al.}(2018)\citenamefont {Li},
  \citenamefont {Wang}, \citenamefont {Mizuseki},\ and\ \citenamefont
  {Chen}}]{Li2018}%
  \BibitemOpen
  \bibfield  {author} {\bibinfo {author} {\bibfnamefont {Z.-Z.}\ \bibnamefont
  {Li}}, \bibinfo {author} {\bibfnamefont {J.-T.}\ \bibnamefont {Wang}},
  \bibinfo {author} {\bibfnamefont {H.}~\bibnamefont {Mizuseki}}, \ and\
  \bibinfo {author} {\bibfnamefont {C.}~\bibnamefont {Chen}},\ }\href {\doibase
  10.1103/PhysRevB.98.094107} {\bibfield  {journal} {\bibinfo  {journal} {Phys.
  Rev. B}\ }\textbf {\bibinfo {volume} {98}},\ \bibinfo {pages} {094107}
  (\bibinfo {year} {2018})}\BibitemShut {NoStop}%
\bibitem [{\citenamefont {Zhu}\ \emph {et~al.}(2020)\citenamefont {Zhu},
  \citenamefont {Yan}, \citenamefont {Liu}, \citenamefont {Oganov},\ and\
  \citenamefont {Zhu}}]{Zhu2020}%
  \BibitemOpen
  \bibfield  {author} {\bibinfo {author} {\bibfnamefont {S.}~\bibnamefont
  {Zhu}}, \bibinfo {author} {\bibfnamefont {X.}~\bibnamefont {Yan}}, \bibinfo
  {author} {\bibfnamefont {J.}~\bibnamefont {Liu}}, \bibinfo {author}
  {\bibfnamefont {A.}~\bibnamefont {Oganov}}, \ and\ \bibinfo {author}
  {\bibfnamefont {Q.}~\bibnamefont {Zhu}},\ }\href {\doibase
  10.2139/ssrn.3565046} {\bibfield  {journal} {\bibinfo  {journal} {SSRN
  Electron. J.}\ ,\ \bibinfo {pages} {1}} (\bibinfo {year} {2020})}\BibitemShut
  {NoStop}%
\bibitem [{\citenamefont {Dong}\ \emph {et~al.}(2020)\citenamefont {Dong},
  \citenamefont {Yao}, \citenamefont {Yao}, \citenamefont {Li}, \citenamefont
  {Hu}, \citenamefont {Zhu}, \citenamefont {Wang}, \citenamefont {Sun},
  \citenamefont {Sundqvist}, \citenamefont {Yang},\ and\ \citenamefont
  {Liu}}]{Dong2020}%
  \BibitemOpen
  \bibfield  {author} {\bibinfo {author} {\bibfnamefont {J.}~\bibnamefont
  {Dong}}, \bibinfo {author} {\bibfnamefont {Z.}~\bibnamefont {Yao}}, \bibinfo
  {author} {\bibfnamefont {M.}~\bibnamefont {Yao}}, \bibinfo {author}
  {\bibfnamefont {R.}~\bibnamefont {Li}}, \bibinfo {author} {\bibfnamefont
  {K.}~\bibnamefont {Hu}}, \bibinfo {author} {\bibfnamefont {L.}~\bibnamefont
  {Zhu}}, \bibinfo {author} {\bibfnamefont {Y.}~\bibnamefont {Wang}}, \bibinfo
  {author} {\bibfnamefont {H.}~\bibnamefont {Sun}}, \bibinfo {author}
  {\bibfnamefont {B.}~\bibnamefont {Sundqvist}}, \bibinfo {author}
  {\bibfnamefont {K.}~\bibnamefont {Yang}}, \ and\ \bibinfo {author}
  {\bibfnamefont {B.}~\bibnamefont {Liu}},\ }\href {\doibase
  10.1103/PhysRevLett.124.065701} {\bibfield  {journal} {\bibinfo  {journal}
  {Phys. Rev. Lett.}\ }\textbf {\bibinfo {volume} {124}},\ \bibinfo {pages}
  {65701} (\bibinfo {year} {2020})}\BibitemShut {NoStop}%
\bibitem [{\citenamefont {Baek}\ \emph {et~al.}(2019)\citenamefont {Baek},
  \citenamefont {Gromilov}, \citenamefont {Kuklin}, \citenamefont {Kovaleva},
  \citenamefont {Fedorov}, \citenamefont {Sukhikh}, \citenamefont {Hanfland},
  \citenamefont {Pomogaev}, \citenamefont {Melchakova}, \citenamefont
  {Avramov},\ and\ \citenamefont {Yusenko}}]{Baek2019}%
  \BibitemOpen
  \bibfield  {author} {\bibinfo {author} {\bibfnamefont {W.}~\bibnamefont
  {Baek}}, \bibinfo {author} {\bibfnamefont {S.~A.}\ \bibnamefont {Gromilov}},
  \bibinfo {author} {\bibfnamefont {A.~V.}\ \bibnamefont {Kuklin}}, \bibinfo
  {author} {\bibfnamefont {E.~A.}\ \bibnamefont {Kovaleva}}, \bibinfo {author}
  {\bibfnamefont {A.~S.}\ \bibnamefont {Fedorov}}, \bibinfo {author}
  {\bibfnamefont {A.~S.}\ \bibnamefont {Sukhikh}}, \bibinfo {author}
  {\bibfnamefont {M.}~\bibnamefont {Hanfland}}, \bibinfo {author}
  {\bibfnamefont {V.~A.}\ \bibnamefont {Pomogaev}}, \bibinfo {author}
  {\bibfnamefont {I.~A.}\ \bibnamefont {Melchakova}}, \bibinfo {author}
  {\bibfnamefont {P.~V.}\ \bibnamefont {Avramov}}, \ and\ \bibinfo {author}
  {\bibfnamefont {K.~V.}\ \bibnamefont {Yusenko}},\ }\href {\doibase
  10.1021/acs.nanolett.8b04421} {\bibfield  {journal} {\bibinfo  {journal}
  {Nano Lett.}\ }\textbf {\bibinfo {volume} {19}},\ \bibinfo {pages} {1570}
  (\bibinfo {year} {2019})}\BibitemShut {NoStop}%
\bibitem [{\citenamefont {He}\ \emph {et~al.}(2002)\citenamefont {He},
  \citenamefont {Sekine},\ and\ \citenamefont {Kobayashi}}]{He2002}%
  \BibitemOpen
  \bibfield  {author} {\bibinfo {author} {\bibfnamefont {H.}~\bibnamefont
  {He}}, \bibinfo {author} {\bibfnamefont {T.}~\bibnamefont {Sekine}}, \ and\
  \bibinfo {author} {\bibfnamefont {T.}~\bibnamefont {Kobayashi}},\ }\href
  {\doibase 10.1063/1.1495078} {\bibfield  {journal} {\bibinfo  {journal}
  {Appl. Phys. Lett.}\ }\textbf {\bibinfo {volume} {81}},\ \bibinfo {pages}
  {610} (\bibinfo {year} {2002})}\BibitemShut {NoStop}%
\bibitem [{\citenamefont {Murri}\ \emph {et~al.}(2019)\citenamefont {Murri},
  \citenamefont {Smith}, \citenamefont {McColl}, \citenamefont {Hart},
  \citenamefont {Alvaro}, \citenamefont {Jones}, \citenamefont {N{\'{e}}meth},
  \citenamefont {Salzmann}, \citenamefont {Cor{\`{a}}}, \citenamefont
  {Domeneghetti}, \citenamefont {Nestola}, \citenamefont {Sobolev},
  \citenamefont {Vishnevsky}, \citenamefont {Logvinova},\ and\ \citenamefont
  {McMillan}}]{Murri2019}%
  \BibitemOpen
  \bibfield  {author} {\bibinfo {author} {\bibfnamefont {M.}~\bibnamefont
  {Murri}}, \bibinfo {author} {\bibfnamefont {R.~L.}\ \bibnamefont {Smith}},
  \bibinfo {author} {\bibfnamefont {K.}~\bibnamefont {McColl}}, \bibinfo
  {author} {\bibfnamefont {M.}~\bibnamefont {Hart}}, \bibinfo {author}
  {\bibfnamefont {M.}~\bibnamefont {Alvaro}}, \bibinfo {author} {\bibfnamefont
  {A.~P.}\ \bibnamefont {Jones}}, \bibinfo {author} {\bibfnamefont
  {P.}~\bibnamefont {N{\'{e}}meth}}, \bibinfo {author} {\bibfnamefont {C.~G.}\
  \bibnamefont {Salzmann}}, \bibinfo {author} {\bibfnamefont {F.}~\bibnamefont
  {Cor{\`{a}}}}, \bibinfo {author} {\bibfnamefont {M.~C.}\ \bibnamefont
  {Domeneghetti}}, \bibinfo {author} {\bibfnamefont {F.}~\bibnamefont
  {Nestola}}, \bibinfo {author} {\bibfnamefont {N.~V.}\ \bibnamefont
  {Sobolev}}, \bibinfo {author} {\bibfnamefont {S.~A.}\ \bibnamefont
  {Vishnevsky}}, \bibinfo {author} {\bibfnamefont {A.~M.}\ \bibnamefont
  {Logvinova}}, \ and\ \bibinfo {author} {\bibfnamefont {P.~F.}\ \bibnamefont
  {McMillan}},\ }\href {\doibase 10.1038/s41598-019-46556-3} {\bibfield
  {journal} {\bibinfo  {journal} {Sci. Rep.}\ }\textbf {\bibinfo {volume}
  {9}},\ \bibinfo {pages} {10334} (\bibinfo {year} {2019})}\BibitemShut
  {NoStop}%
\bibitem [{\citenamefont {Glass}\ \emph {et~al.}(2006)\citenamefont {Glass},
  \citenamefont {Oganov},\ and\ \citenamefont {Hansen}}]{Glass2006}%
  \BibitemOpen
  \bibfield  {author} {\bibinfo {author} {\bibfnamefont {C.~W.}\ \bibnamefont
  {Glass}}, \bibinfo {author} {\bibfnamefont {A.~R.}\ \bibnamefont {Oganov}}, \
  and\ \bibinfo {author} {\bibfnamefont {N.}~\bibnamefont {Hansen}},\ }\href
  {https://linkinghub.elsevier.com/retrieve/pii/S0010465506002931} {\bibfield
  {journal} {\bibinfo  {journal} {Comput. Phys. Commun.}\ }\textbf {\bibinfo
  {volume} {175}},\ \bibinfo {pages} {713} (\bibinfo {year}
  {2006})}\BibitemShut {NoStop}%
\bibitem [{\citenamefont {Oganov}\ and\ \citenamefont
  {Glass}(2006)}]{Oganov2006}%
  \BibitemOpen
  \bibfield  {author} {\bibinfo {author} {\bibfnamefont {A.~R.}\ \bibnamefont
  {Oganov}}\ and\ \bibinfo {author} {\bibfnamefont {C.~W.}\ \bibnamefont
  {Glass}},\ }\href {\doibase 10.1063/1.2210932} {\bibfield  {journal}
  {\bibinfo  {journal} {J. Chem. Phys.}\ }\textbf {\bibinfo {volume} {124}},\
  \bibinfo {pages} {244704} (\bibinfo {year} {2006})}\BibitemShut {NoStop}%
\bibitem [{\citenamefont {Lyakhov}\ \emph {et~al.}(2010)\citenamefont
  {Lyakhov}, \citenamefont {Oganov},\ and\ \citenamefont
  {Valle}}]{Lyakhov2010}%
  \BibitemOpen
  \bibfield  {author} {\bibinfo {author} {\bibfnamefont {A.~O.}\ \bibnamefont
  {Lyakhov}}, \bibinfo {author} {\bibfnamefont {A.~R.}\ \bibnamefont {Oganov}},
  \ and\ \bibinfo {author} {\bibfnamefont {M.}~\bibnamefont {Valle}},\ }\href
  {\doibase 10.1016/j.cpc.2010.06.007} {\bibfield  {journal} {\bibinfo
  {journal} {Comput. Phys. Commun.}\ }\textbf {\bibinfo {volume} {181}},\
  \bibinfo {pages} {1623} (\bibinfo {year} {2010})}\BibitemShut {NoStop}%
\bibitem [{\citenamefont {Jain}\ \emph {et~al.}(2013)\citenamefont {Jain},
  \citenamefont {Ong}, \citenamefont {Hautier}, \citenamefont {Chen},
  \citenamefont {Richards}, \citenamefont {Dacek}, \citenamefont {Cholia},
  \citenamefont {Gunter}, \citenamefont {Skinner}, \citenamefont {Ceder},\ and\
  \citenamefont {Persson}}]{Jain2013}%
  \BibitemOpen
  \bibfield  {author} {\bibinfo {author} {\bibfnamefont {A.}~\bibnamefont
  {Jain}}, \bibinfo {author} {\bibfnamefont {S.~P.}\ \bibnamefont {Ong}},
  \bibinfo {author} {\bibfnamefont {G.}~\bibnamefont {Hautier}}, \bibinfo
  {author} {\bibfnamefont {W.}~\bibnamefont {Chen}}, \bibinfo {author}
  {\bibfnamefont {W.~D.}\ \bibnamefont {Richards}}, \bibinfo {author}
  {\bibfnamefont {S.}~\bibnamefont {Dacek}}, \bibinfo {author} {\bibfnamefont
  {S.}~\bibnamefont {Cholia}}, \bibinfo {author} {\bibfnamefont
  {D.}~\bibnamefont {Gunter}}, \bibinfo {author} {\bibfnamefont
  {D.}~\bibnamefont {Skinner}}, \bibinfo {author} {\bibfnamefont
  {G.}~\bibnamefont {Ceder}}, \ and\ \bibinfo {author} {\bibfnamefont {K.~A.}\
  \bibnamefont {Persson}},\ }\href {\doibase 10.1063/1.4812323} {\bibfield
  {journal} {\bibinfo  {journal} {APL Mater.}\ }\textbf {\bibinfo {volume}
  {1}},\ \bibinfo {pages} {011002} (\bibinfo {year} {2013})}\BibitemShut
  {NoStop}%
\bibitem [{\citenamefont {Gra{\v{z}}ulis}\ \emph {et~al.}(2009)\citenamefont
  {Gra{\v{z}}ulis}, \citenamefont {Chateigner}, \citenamefont {Downs},
  \citenamefont {Yokochi}, \citenamefont {Quir{\'{o}}s}, \citenamefont
  {Lutterotti}, \citenamefont {Manakova}, \citenamefont {Butkus}, \citenamefont
  {Moeck},\ and\ \citenamefont {{Le Bail}}}]{Grazulis2009}%
  \BibitemOpen
  \bibfield  {author} {\bibinfo {author} {\bibfnamefont {S.}~\bibnamefont
  {Gra{\v{z}}ulis}}, \bibinfo {author} {\bibfnamefont {D.}~\bibnamefont
  {Chateigner}}, \bibinfo {author} {\bibfnamefont {R.~T.}\ \bibnamefont
  {Downs}}, \bibinfo {author} {\bibfnamefont {A.~F.~T.}\ \bibnamefont
  {Yokochi}}, \bibinfo {author} {\bibfnamefont {M.}~\bibnamefont
  {Quir{\'{o}}s}}, \bibinfo {author} {\bibfnamefont {L.}~\bibnamefont
  {Lutterotti}}, \bibinfo {author} {\bibfnamefont {E.}~\bibnamefont
  {Manakova}}, \bibinfo {author} {\bibfnamefont {J.}~\bibnamefont {Butkus}},
  \bibinfo {author} {\bibfnamefont {P.}~\bibnamefont {Moeck}}, \ and\ \bibinfo
  {author} {\bibfnamefont {A.}~\bibnamefont {{Le Bail}}},\ }\href {\doibase
  10.1107/S0021889809016690} {\bibfield  {journal} {\bibinfo  {journal} {J.
  Appl. Crystallogr.}\ }\textbf {\bibinfo {volume} {42}},\ \bibinfo {pages}
  {726} (\bibinfo {year} {2009})}\BibitemShut {NoStop}%
\bibitem [{\citenamefont {Kresse}\ and\ \citenamefont
  {Furthm{\"{u}}ller}(1996)}]{Kresse:1996}%
  \BibitemOpen
  \bibfield  {author} {\bibinfo {author} {\bibfnamefont {G.}~\bibnamefont
  {Kresse}}\ and\ \bibinfo {author} {\bibfnamefont {J.}~\bibnamefont
  {Furthm{\"{u}}ller}},\ }\href {\doibase 10.1103/PhysRevB.54.11169} {\bibfield
   {journal} {\bibinfo  {journal} {Phys. Rev. B}\ }\textbf {\bibinfo {volume}
  {54}},\ \bibinfo {pages} {11169} (\bibinfo {year} {1996})}\BibitemShut
  {NoStop}%
\end{thebibliography}%
\pagebreak{}

\section*{Supplemental Figures}

\begin{figure}
\includegraphics[scale=0.3]{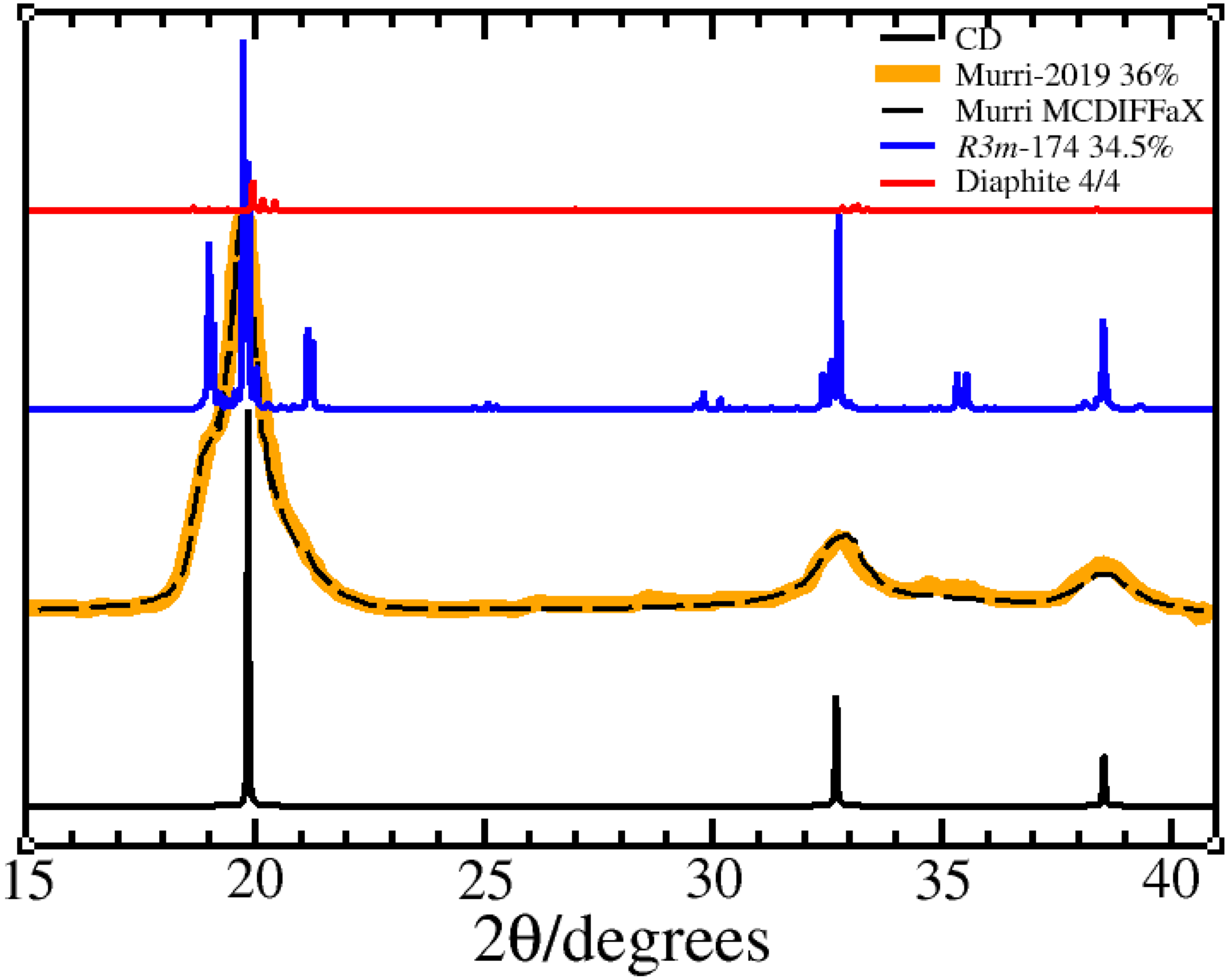}

\caption{\label{fig:Diaphite-XRD}}
\end{figure}

\end{document}